\documentstyle[aps,multicol,amsfonts,amssymb,amsmath,graphicx,axodraw]{revtex}

\makeatletter
\@addtoreset{equation}{section}
\makeatother
\newcommand{\tr}{{\mathrm tr}}

\begin{document}

\draft
\title{\LARGE \bf On the Chiral Phase Transition\\ in the Linear Sigma Model}
\author{Tran Huu Phat$^{1)}$, Nguyen Tuan Anh$^{2)}$ and Le Viet Hoa$^{3)}$}
\address{$^{1)}$ Vietnam Atomic Energy Commission, \\ 59 Ly Thuong Kiet, Hanoi, Vietnam.\\
$^{2)}$ Institute for Nuclear Science and Technique,\\ P.O.Box 5T-160, Hoang Quoc Viet, Hanoi, Vietnam \\
$^{3)}$ Department of Physics, Hanoi University of Education II,
 Hanoi, Vietnam. }
\date{July 2003}
\maketitle

\begin{abstract} {\normalsize
The Cornwall-Jackiw-Tomboulis (CJT) effective action for composite
operators at finite temperature is used to investigate the chiral
phase transition within the framework of the linear sigma model as
the low-energy effective model of quantum chromodynamics (QCD). A
new renormalization prescription for the CJT effective action in
the Hartree-Fock (HF) approximation is proposed. A numerical
study, which incorporates both thermal and quantum effect, shows
that in this approximation the phase transition is of first order.
However, taking into account the higher-loop diagrams contribution
the order of phase transition is unchanged.}
\end{abstract}




\large

\section{\large \bf INTRODUCTION}

In recent years the study of the phase transitions has become the
subject of intense investigation since it is very important from
various aspects, ranging from cosmology to particle physics.
According to the big bang model there is a series of phase
transitions, including, of course, the chiral phase one, at the
early stage of the universe evolution. The nature of the
electroweak phase transition has called much attention due to the
suggestion that the baryon asymmetry may be generated at the
electroweak scale if the transition is of first order
\cite{hutp91}. The restoration of chiral symmetry may lead to
several interesting phenomena observed, for instance, in the
dilepton mass spectrum \cite{npa638} or in the formation of
disoriented chiral condensates \cite{raj95}. The quantum
chromodynamics (QCD) is chirally symmetric for light quark sector
and must be spontaneously broken due to the existence of pions,
the Nambu-Goldstone (NG) bosons. The lattice QCD calculations
indicate that the restoration of chiral symmetry occurs at
temperatures of order 150 MeV \cite{npb53}, which is expected to
be probed in ultrarelativistic heavy-ion collision experiments
planned at RHIC and LHC \cite{npa610}. The field theoretical
investigation of the symmetry restoration at finite temperatures
was first carried out by Kirzhnits and Linde \cite{plb42} and then
was systematically developed by Weinberg \cite{prd9a}, Dolan and
Jackiw \cite{prd9b} as well as many others adopting the effective
action as the most appropriate formalism for such studies. It was
proved that the leading contributions at very high temperatures
come from infinite series of certain class of multiloop diagrams
in perturbation theory, which, in the $\lambda \phi^4$ theory, are
the so-called daisy and superdaisy diagrams. Recently the
effective action for composite operators, originally introduced by
Cornwall, Jackiw and Tomboulis \cite{prd10} at zero temperature,
has been extended for finite temperatures in $\lambda \phi^4$
theory by Amelino-Camelia and Pi \cite{prd47a}, who showed that
one needs to resume only ``double bubble'' graphs instead of
summing infinite set of daisy and superdaisy graphs using the tree
level propagators. This is known as Hartree-Fock approximation
\cite{plb201}. Moreover, the CJT formalism at finite temperatures
leads to reliable results [12, 13], among other approximate
approachs.

Many works on chiral symmetry restoration at high temperatures
have been accomplished [14-20] within the framework of the linear
sigma model, which is considered to be the relevant model
\cite{prd37} for effective theory in low-energy phenomenology of
QCD modelling the hadron dynamics associated to the chiral phase
transition. However, there exists serious difficulty concerning
the renormalization of the CJT effective action in the
Hartree-Fock approximation. The point is that a consistent
renormalization cannot be performed in the broken phase because
all terms in the effective action are expressed not only by
renormalized quantities but bare quantities remain in several
terms \cite{hepph97}. In [16,19] one tried to regularize divergent
integrals by means of several schemes instead of renormalizing the
CJT effective action in Hartree-Fock approximation at finite
temperatures.

In this work, dealing with the chiral phase transition at finite
temperatures, we re-examine the linear sigma model since it is
most suited for our purpose of describing the phase transition in
a self-consistent approximation. We do not consider the general
$O(N)$ linear sigma model at large-N limit because it
significantly simplifies the counting of multiloop bubble
diagrams, which is the most common source of inaccuracy.

We use the imaginary time formalism of Matsubara \cite{kap89} and
work in Euclidean space-time. The Feynman rules are the same as
those at zero temperature, except that
\[
\int\frac{d^4k}{(2\pi)^4} f(k) \rightarrow \frac{1}{\beta}\sum_n
\int\frac{d^3k}{(2\pi)^3} f(\omega_n,\vec{k}) \equiv \int_\beta
f(k),
\]
where $\omega_n = 2\pi n/\beta$.

This paper is organized as follows. Section~\ref{sec:CJTp} is
devoted to calculating the CJT effective potential at finite
temperatures in the Hartree-Fock approximation. Here a new
renormalization prescription is proposed. A numerical computation
is presented in Section~\ref{sec:HFnum}. The higher-loop diagram
contribution in the vicinity of critical temperature is considered
in Section~\ref{sec:Hloop}. Finally conclusion and discussion are
given in Section~\ref{sec:Conc}.

\section{\large \bf THE CJT EFFECTIVE POTENTIAL}
\label{sec:CJTp}

The lagrangian density of the $O(4)$ linear sigma model reads
\begin{eqnarray}
{\cal L} = \frac{1}{2} (\partial_\mu \phi^\alpha)^2 +\frac{m^2}2
\phi^2 +\frac{\lambda}{24} (\phi^2)^2, \label{2.1}
\end{eqnarray}
where $\phi^2 = \phi^\alpha\phi^\alpha$; $\phi^i = \pi^i$,
$i=1,2,3$; $\phi^4 = \sigma$.

The counterterms, which must be added to (\ref{2.1}), are chosen
as
\begin{eqnarray}
\Delta {\cal L}=\frac{\delta m^2}2 \phi^2 +\frac{\delta
\lambda}{24} (\phi^2)^2 +\delta\Omega, \label{2.2}
\end{eqnarray}
in which the counterterm $\delta\Omega$ is introduced for vacuum
energy. As a rule, all divergencies of the theory have to be
absorbed by the counterterms (\ref{2.2}).

By shifting the field as $\phi^\alpha(x) \rightarrow
\phi^\alpha(x) +\phi_0^\alpha$, the tree-level propagators
$D_{\alpha\beta}$ have the form
\begin{eqnarray}
D_{\alpha\beta}^{-1} &=& \left.\frac{\delta^2
I(\phi)}{\delta\phi^\alpha(x)\delta\phi^\beta(y)}\right|_{\phi=\phi_0}
\nonumber \\ &=&
 \left[ \partial_\mu^2 +m^2+\delta
m^2+\frac{\lambda+\delta\lambda}6 \phi_0^2\right]
\delta^{\alpha\beta}\delta^4(x-y)+\frac{\lambda+\delta\lambda}3\phi_0^\alpha\phi_0^\beta
\delta^4(x-y). \label{2.3}
\end{eqnarray}
Here $I(\phi)$ is the classical action
\[
I(\phi) = \int d^4x ({\cal L} +\Delta{\cal L}).
\]
In momentum space, $D_{\alpha\beta}^{-1}$ is written as
\begin{eqnarray}
D_{\alpha\beta}^{-1}(k;\phi_0) = \left[ k^2 +m^2+\delta
m^2+\frac{\lambda+\delta\lambda}6 \phi_0^2\right]
\delta^{\alpha\beta} + \frac{\lambda+\delta\lambda}3
\phi_0^\alpha\phi_0^\beta. \label{2.4}
\end{eqnarray}

\begin{figure}[h]
\begin{center}
\begin{picture}(350,20)(0,0)
\GCirc(20,0){20}{1} \GCirc(60,0){20}{1}  \Vertex(40,0){1.5}
\GCirc(120,0){20}{1} \Line(120,20)(120,-20) \Vertex(120,20){1.5}
\Vertex(120,-20){1.5} \GCirc(180,0){20}{1} \CArc(160,0)(29,-45,45)
\CArc(200,0)(29,135,225) \Vertex(180,20){1.5}
\Vertex(180,-20){1.5} \GCirc(240,0){20}{1} \Line(223,10)(240,-20)
\Line(257,10)(240,-20) \Vertex(223,10){1.5} \Vertex(240,-20){1.5}
\Vertex(257,10){1.5} \GCirc(300,0){20}{1} \Line(300,20)(300,-20)
\Line(300,0)(320,0) \Vertex(300,20){1.5} \Vertex(300,-20){1.5}
\Vertex(300,0){1.5} \Vertex(320,0){1.5} \Text(340,0)[l]{$\cdots $}
\end{picture} \vspace{1cm}
\caption{Two-particle irreducible graphs contributing to
$\Gamma_2(\phi,G)$ up to the three-loop level in a $\lambda
\phi^4$ theory. The solid line represents the propagator $G(x,y)$.
There are two kinds of vertices: a three-particle vertex
proportional to $\lambda\phi$ and a four-particle vertex.
\label{fig:Fig0}}
\end{center}
\end{figure}
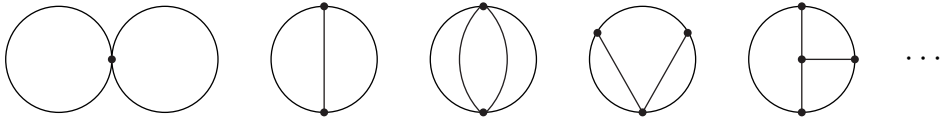

The interaction action, describing the vertices of the shifted
theory, is given by
\[
I_{int}(\phi,\phi_0) =\int d^4x \left[
\frac{\lambda+\delta\lambda}6 \phi^2 \phi^\alpha\phi_0^\alpha
+\frac{\lambda+\delta\lambda}{24}(\phi^2)^2\right].
\]
Then, for constant $\phi_0$, we arrive at the CJT effective
potential at finite temperatures
\begin{eqnarray*}
V_\beta^{CJT}(\phi_0,G) = I(\phi_0) +\frac{1}2 \int_\beta \tr\ln
G^{-1}(k) +\frac{1}2 \int_\beta \tr\left[ D^{-1}(k;\phi_0)G(k)
\right] +V_2^{CJT}(\phi_0,G).
\end{eqnarray*}
Here $G_{\alpha\beta}(k)$ is the full propagator of the theory and
$V_2^{CJT}$ is the sum of all two and higher order loop
two-particle irreducible vacuum graphs of the theory with vertices
given by $I_{int}(\phi,\phi_0)$ and propagators set equal to
$G_{\alpha\beta}(k)$. The graphs shown in Fig.~\ref{fig:Fig0} are
under discussion. It is clear that, among them, only the two-loop
graph, the ``double bubble'' one, of order $O(\lambda)$ includes
contributions from daisy and superdaisy graphs of ordinary
perturbation theory. It is known that truncating the series at
$O(\lambda)$ is the Hartree-Fock approximation.

Therefore, the CJT effective potential at finite temperatures in
the Hartree-Fock approximation is derived
\begin{eqnarray}
V_\beta^{CJT}(\phi_0,G) &=& \delta\Omega +\frac{m^2+\delta m^2}2
\phi_0^2 +\frac{\lambda+\delta\lambda}{24}(\phi_0^2)^2 \nonumber
\\ & & +\frac{1}2 \int_\beta\tr\ln G^{-1}(k) +\frac{1}2 \int_\beta
\tr\left[D^{-1}(k;\phi_0)G \right] \nonumber \\ & &
+\frac{\lambda+\delta\lambda}{24}\left\{ \left[\int_\beta
G_{\alpha\alpha}(k)\right]^2 +2 \int_\beta
G_{\alpha\beta}(k)\int_\beta G_{\beta\alpha}(k)\right\}.
\label{2.5}
\end{eqnarray}
Minimizing $V_\beta^{CJT}(\phi_0,G)$ with respect to
$G_{\alpha\beta}(k)$ we obtain the Schwinger-Dyson (SD) equations
for propagators
\begin{eqnarray}
G_{\alpha\beta}^{-1}(p) = D_{\alpha\beta}^{-1}(p;\phi_0) +
\frac{\lambda+\delta\lambda}6 \left[
\delta_{\alpha\beta}\int_\beta G_{\delta\delta}(k) + 2\int_\beta
G_{\alpha\beta}(k)\right]. \label{2.6}
\end{eqnarray}
Basing on the structure of (\ref{2.6}) and the $O(4)$ symmetry we
adopt the following ansatz for propagators $G_{\alpha\beta}$ in
Hartree-Fock approximation
\begin{eqnarray}
G_{\alpha\beta}^{-1}(p) = (p^2 + X^2)\delta_{\alpha\beta} +Y^2
\frac{\phi_0^\alpha \phi_0^\beta}{\phi_0^2}. \label{2.7}
\end{eqnarray}
Inserting (\ref{2.7}) into (\ref{2.5}) provides
\begin{eqnarray}
V_\beta^{CJT}(\phi_0,M_\sigma,M_\pi) &=& \delta\Omega
+\frac{m^2+\delta m^2}2 \phi_0^2
+\frac{\lambda+\delta\lambda}{24}(\phi_0^2)^2 +Q(M_\sigma)
+3Q(M_\pi) \nonumber \\  & +&\frac{1}2 \left( m^2 +\delta
m^2+\frac{\lambda+\delta\lambda}2 \phi_0^2 -M_\sigma^2\right)
P(M_\sigma) \nonumber \\  & +& \frac{3}2 \left( m^2 +\delta
m^2+\frac{\lambda+\delta\lambda}6 \phi_0^2 -M_\pi^2\right)
P(M_\pi) \nonumber \\  & +&\frac{\lambda+\delta\lambda}{24}
\left\{ 3 \left[ P(M_\sigma)\right]^2 +15 \left[ P(M_\pi)\right]^2
+ 6 P(M_\sigma)P(M_\pi) \right\}, \label{2.8}
\end{eqnarray}
where
\begin{eqnarray*}
M_\sigma &=& X^2 +Y^2, \\
M_\pi &=& X^2,
\end{eqnarray*}
and
\begin{eqnarray*}
Q(M) &=& \frac{1}2 \int_\beta \ln (k^2 +M^2), \\
P(M) &=& \int_\beta \frac{1}{k^2+M^2}.
\end{eqnarray*}

In order to regularize the divergent integrals $P(M)$ and $Q(M)$,
contained in the expression (\ref{2.8}) of the effective
potential, let us use the three-dimensional momentum cutoff
scheme, in which every divergent integral is written as the sum of
a divergent part and a finite part, namely
\begin{eqnarray*}
Q(M) &=& {\mathrm Div} Q(M) + Q_f(M), \\
{\mathrm Div} Q(M) &=& -\frac{M^4}4 I_2 +\frac{M^2}2 I_1, \\
Q_f(M) &=& \frac{M^4}{64\pi^2}
\left(\ln\frac{M^2}{\mu^2}-\frac{1}2 \right)
+T\int\frac{d^3k}{(2\pi)^3}\ln\left(1-e^{-\frac{E(\vec{k})}T}\right),
\end{eqnarray*}
\begin{eqnarray*}
P(M) &=& {\mathrm Div} P(M) + P_f(M), \\
{\mathrm Div} P(M) &=& I_1 - M^2 I_2, \\
P_f(M) &=& \frac{M^2}{16\pi^2} \ln\frac{M^2}{\mu^2}
-\int\frac{d^3k}{(2\pi)^3}\left[
E(\vec{k})\left(1-e^{\frac{E(\vec{k})}T}\right)\right]^{-1},
\end{eqnarray*}
\[
I_1 = \frac{\Lambda^2}{8\pi^2}, \;\;\; I_2 =
\frac{1}{16\pi^2}\ln\frac{\Lambda^2}{\mu^2}, \;\;\; E(\vec{k}) =
(\vec{k}^2 +M^2)^{1/2}.
\]

Now the renormalization is carried out so that all infinite terms
appearing in (\ref{2.8}) must be eliminated. For this end, the
counterterms $\delta\Omega$, $\delta m^2$ and $\delta\lambda$ are
chosen to obey the following constraints:
\begin{eqnarray}
& & \delta\Omega +{\mathrm Div} Q(M_\sigma) +3{\mathrm Div}
Q(M_\pi)=0, \label{2.9} \\
& & \frac{\delta m^2}2 \left[ \phi_0^2 +P(M_\sigma)
+3P(M_\pi)\right] \nonumber \\ & &+\frac{\delta\lambda}4 \left\{
\frac{(\phi_0^2)^2}6 +\phi_0^2 P(M_\sigma) +\phi_0^2 P(M_\pi)
+\frac{1}2 \left[ P(M_\sigma)\right]^2 +\frac{5}2 \left[
P(M_\pi)\right]^2 + P(M_\sigma)P(M_\pi)\right\} \nonumber \\
& & +\frac{1}2 \left( m^2 +\frac{\lambda}2 \phi_0^2
-M_\sigma^2\right) {\mathrm Div}P(M_\sigma) + \frac{3}2 \left( m^2
+\frac{\lambda}6 \phi_0^2 -M_\pi^2\right) {\mathrm Div}P(M_\pi)
\nonumber \\ & & +\frac{\lambda}{24} \left\{ 3 {\mathrm Div}\left[
P(M_\sigma)\right]^2 + 15 {\mathrm Div}\left[ P(M_\pi)\right]^2 +
6 {\mathrm Div}\left[ P(M_\sigma)P(M_\pi)\right]\right\} = 0,
\label{2.10}
\end{eqnarray}
(\ref{2.10}) is a linear equation of two unknowns $\delta m^2$ and
$\delta \lambda$, it has an infinite number of roots. In the
following are two simplest cases:

$a)$ $\delta m^2 =0$,

$b)$ $\delta \lambda =0$.

In the first case $\delta\lambda$ tends to a finite limit as
$\Lambda \rightarrow +\infty$ and in the second case $\delta m^2$
tends to an infinity as $\Lambda \rightarrow +\infty$. As a rule,
both cases are equally accepted for renormalizing the CJT
effective potential in Hartree-Fock approximation. In general, the
counterterms are temperature dependent as was mentioned in
\cite{prd15}.

We obtain ultimately the renormalized effective potential at
finite temperatures
\begin{eqnarray}
V_\beta^{CJT}(\phi_0,M_\sigma,M_\pi) &=& \frac{m^2}2 \phi_0^2
+\frac{\lambda}{24}(\phi_0^2)^2 +Q_f(M_\sigma) +3Q_f(M_\pi)
\nonumber \\ & & +\frac{1}2 \left(m^2 +\frac{\lambda}2 \phi_0^2
-M^2_\sigma\right)P_f(M_\sigma) +\frac{3}2 \left(m^2
+\frac{\lambda}6 \phi_0^2 -M^2_\pi\right)P_f(M_\pi) \nonumber \\ &
& +\frac{\lambda}8 \left\{ \left[ P_f(M_\sigma)\right]^2 + 5\left[
P_f(M_\pi)\right]^2 + 2 P_f(M_\sigma)P_f(M_\pi) \right\}.
\label{2.11}
\end{eqnarray}
The SD equations for $M_\sigma^2$ and $M_\pi^2$ and the gap
equation for the sigma condensate are, respectively, derived from
(\ref{2.11}) for $\phi_0^4 =\sigma$, $\phi_0^i = \pi^i = 0$,
\begin{eqnarray}
M_\sigma^2 &=& m^2 +\frac{\lambda}2 \sigma_0^2 +\frac{\lambda}2
P_f(M_\sigma) +\frac{\lambda}2 P_f(M_\pi), \label{2.12} \\ M_\pi^2
&=& m^2 +\frac{\lambda}6 \sigma_0^2 +\frac{\lambda}6 P_f(M_\sigma)
+\frac{5\lambda}6 P_f(M_\pi), \label{2.13}
\end{eqnarray}
and
\[
\left( \frac{d V_\beta^{CJT}}{d\phi_0^4}\right)_{\begin{array}{l}
\phi_0^4 =\sigma_0 \\ \phi_0^i =0 \end{array}} = \sigma_0 \left(
m^2 +\frac{\lambda}6 \sigma_0^2 +\frac{\lambda}2 P_f(M_\sigma)
+\frac{\lambda}2 P(M_\pi) \right) = 0,
\]
or
\begin{eqnarray}
m^2 + \frac{\lambda}6 \sigma_0^2 +\frac{\lambda}2 P_f(M_\sigma)
+\frac{\lambda}2 P(M_\pi) = 0. \label{2.14}
\end{eqnarray}
Taking into account (\ref{2.12}) the equation (\ref{2.14}) reduces
to
\begin{eqnarray}
\frac{\lambda}3 \sigma_0^2 - M_\sigma^2 = 0. \label{2.15}
\end{eqnarray}

Adopting the argument of \cite{epja9} we distinguish the effective
masses of mesons, $M_\sigma$ and $M_\pi$, as those appearing in
the approximate propagators, from the physical masses which are
defined as the curvatures of the effective potential around its
minimum,
\begin{eqnarray}
m_\sigma^2 &\overset{\mbox{def.}}{=}& \left( \frac{d^2
V_\beta^{CJT}}{d\sigma^2}\right)_{\begin{array}{l} \sigma
=\sigma_0 \\ \pi =0 \end{array} }, \label{2.16} \\
m_\pi^2 &\overset{\mbox{def.}}{=}& \left( \frac{d^2
V_\beta^{CJT}}{d\pi^2}\right)_{\begin{array}{l} \sigma =\sigma_0
\\ \pi =0 \end{array} } =0, \label{2.17}
\end{eqnarray}
in the broken symmetry phase.

Equations (\ref{2.12})-(\ref{2.15}) enable us to determine the
``transition temperature'' $T_{c1}$, at which the broken chiral
symmetry is restored: $\sigma_0$, $M_\sigma$ and $M_\pi$ tend to
zero as $T \rightarrow T_{c1} -0$. Indeed, Eq.~(\ref{2.15}) tells
that $M_\sigma \rightarrow 0$ as $\sigma_0 \rightarrow 0$ and
Eq.~(\ref{2.13}) simplifies to
\begin{eqnarray}
M_\pi^2 = m^2 +\lambda P_f(M_\pi). \label{2.18}
\end{eqnarray}
When $M_\pi$ vanishes as $T \rightarrow T_{c1}$, {\it i.e.}
$M_\pi/T_{c1} \ll 1$, we have the expression
\begin{eqnarray}
P_f(M_\pi) \approx T_{c1}^2 \left[ \frac{1}{12}
-\frac{1}{4\pi}\frac{M_\pi}{T_{c1}} -
\frac{1}{16\pi^2}\frac{M_\pi^2}{T^2_{c1}}\ln\frac{M_\pi^2}{T^2_{c1}}\right].
\label{2.19}
\end{eqnarray}
Combining (\ref{2.18}) and (\ref{2.19}) leads to
\[
T_{c1} \approx \sqrt{2} \left( -\frac{6 m^2}{\lambda}\right)^{1/2}
= \sqrt{2}f_\pi \approx 131.5 \mbox{ MeV}.
\]
Here $f_\pi = 93$ MeV is the pion decay constant.

\section{\large \bf NUMERICAL STUDY IN HARTREE-FOCK APPROXIMATION}
\label{sec:HFnum}

Let us first introduce a term explicitly breaking chiral symmetry
into the lagrangian density
\begin{eqnarray}
{\cal L} = \frac{1}2 \left( \partial_\mu\phi^\alpha\right)^2
+\frac{m^2+\delta m^2}2 \phi^2 +\frac{\lambda+\delta\lambda}{24}
(\phi^2)^2 +\delta\Omega -c\sigma. \label{3.1}
\end{eqnarray}
The renomalized effective potential corresponding to (\ref{3.1})
reads
\begin{eqnarray}
V_\beta^{CJT}(\sigma_0,M_\sigma,M_\pi) &=& \frac{m^2}2 \sigma_0^2
+\frac{\lambda}{24}\sigma_0^4 -c\sigma_0 + Q_f(M_\sigma)
+3Q_f(M_\pi) \nonumber
\\ & & +\frac{1}2 \left(m^2 +\frac{\lambda}2 \sigma_0^2
-M_\sigma^2\right)P_f(M_\sigma) +\frac{3}2 \left(m^2
+\frac{\lambda}6 \sigma_0^2 -M_\pi^2\right)P_f(M_\pi) \nonumber \\
& & +\frac{\lambda}8 \left\{ \left[ P_f(M_\sigma)\right]^2 +
5\left[ P_f(M_\pi)\right]^2 + 2 P_f(M_\sigma)P_f(M_\pi) \right\},
\label{3.2}
\end{eqnarray}
which leads to the gap equation for the sigma condensate
$\sigma_0$
\begin{eqnarray}
\sigma_0 \left( M_\sigma^2 -\frac{\lambda}3 \sigma_0^2 \right) =
c. \label{3.3}
\end{eqnarray}

The SD equations for $M_\sigma$ and $M_\pi$, derived from
(\ref{3.2}), have the same form (\ref{2.12}) and (\ref{2.13}).
Inserting these equations into (\ref{3.2}) we arrive at
\begin{eqnarray}
V(\sigma_0,T) &=& \frac{m^2}2\sigma^2_0
+\frac{\lambda}{24}\sigma_0^4 -c\sigma_0 + Q_f(M_\sigma)
+3Q_f(M_\pi) \nonumber \\ & & -\frac{\lambda}8 \left[
P_f(M_\sigma)\right]^2 -\frac{5\lambda}8\left[ P_f(M_\pi)\right]^2
-\frac{\lambda}4 P_f(M_\sigma)P_f(M_\pi). \label{3.4}
\end{eqnarray}

Now the physical mass of pion no longer vanishes in broken
symmetry phase and equals to
\begin{eqnarray}
m_\pi^2 = \frac{c}{\sigma_0}. \label{3.5}
\end{eqnarray}

For numerical computation we use the model parameters at zero
temperature as initial condition, namely
\begin{eqnarray*}
& & c= f_\pi m_\pi^2 \hspace{1cm} \mbox{at } T =0 \\
& & \lambda = \frac{3}{f_\pi^2}(m_\sigma^2 -m_\pi^2 ) \hspace{1cm}
\mbox{at } T
= 0 \\
& & m^2 = -\frac{1}2 m_\sigma^2 +\frac{3}2 m_\pi^2 \hspace{.6cm}
\mbox{at } T = 0.
\end{eqnarray*}
We take $m_\pi = 138$ MeV, $m_\sigma = 500$ MeV and $f_\pi = 93$
MeV at $T=0$.

For convenience, let us consider two separate cases.

\subsection{\large \bf Hartree-Fock approximation in chiral limit $c=0$}

In addition to the preceding parameters, the regularization
introduced another parameter $\mu^2$, which corresponds to
renormalization scale. Therefore, we must first determine a
suitable value $\mu_0^2$ for $\mu^2$, then all other quantities
are determined at $\mu^2 =\mu_0^2$. The value $\mu_0^2$ is
determined as the real root of the following equation
\[
\sigma_0 (\mu^2,0)|_{\mu^2 =\mu_0^2} = f_\pi = 93 \mbox{ MeV},
\]
where $\sigma_0 (\mu^2,0)$ is a solution of the system of
equations (\ref{2.12}), (\ref{2.13}) and (\ref{2.14}) at $T=0$.

The numerical computation gives $\mu_0 = 539.6$ MeV\footnote{which
corresponds to $M_\sigma =481.58$ MeV, $M_\pi = 73.53$ MeV at
$T=0$.}. Eliminating $\sigma_0$ from (\ref{2.12})-(\ref{2.14}) we
get
\begin{eqnarray}
M_\sigma^2 &=& -2m^2 -\lambda P_f(M_\sigma) -\lambda P_f(M_\pi),
\nonumber \\
M_\pi^2 &=& -\frac{\lambda}3 P_f(M_\sigma) +\frac{\lambda}3
P_f(M_\pi). \label{3.6}
\end{eqnarray}

\begin{multicols}{2}
\begin{figure}[h]
\begin{center}
\includegraphics[width=1\columnwidth]{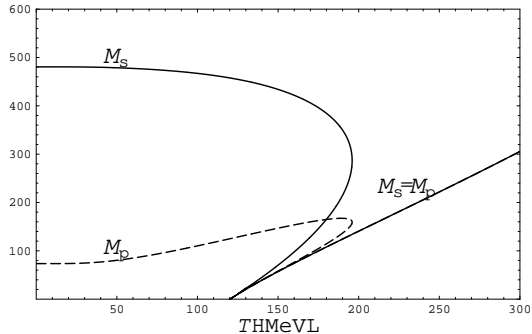}
\vskip .5cm \caption{The temperature dependent solution of the SD
equations in the chiral limit. \label{fig:Fig1}}
\end{center} 
\end{figure}
\begin{figure}[h]
\begin{center}
\includegraphics[width=1\columnwidth]{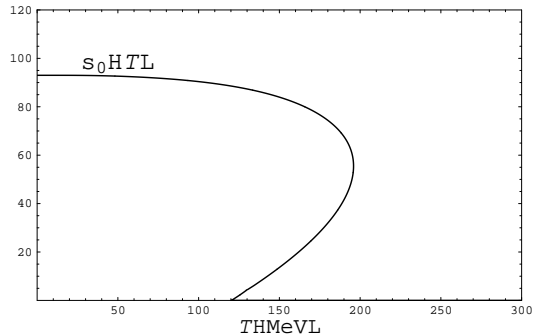}
\vskip .5cm \caption{Evolution of the order parameter as a
function of temperature. \label{fig:Fig2}}
\end{center} 
\end{figure}
\end{multicols}
Inserting $\mu=\mu_0$ into the system (\ref{3.6}) and then solving
numerically the preceding system of two equations we obtain the
solution presented in Fig.~\ref{fig:Fig1}. As is shown in
Fig.~\ref{fig:Fig2} the temperature $T_{c1}$, calculated above by
means of the high temperature approximation, is very close to the
one, obtained by numerically solving the system
(\ref{2.12})-(\ref{2.14}). As the temperature increases from zero,
the order parameter decreases from $f_\pi$, jumps to zero at
$T_{c1}$ and remains zero above $T_{c1}$. In the meanwhile, the
effective masses of mesons, $M_\sigma$ and $M_\pi$, change along
the upper lines, jump to lower coinciding line at $T_{c1}$ and
increase along this line. It is clear that the phase transition is
of first order. The indication of a first order phase transition
has been reported in many publications [14-20].

\begin{figure}[h]
\begin{center}
\includegraphics[width=.6\columnwidth]{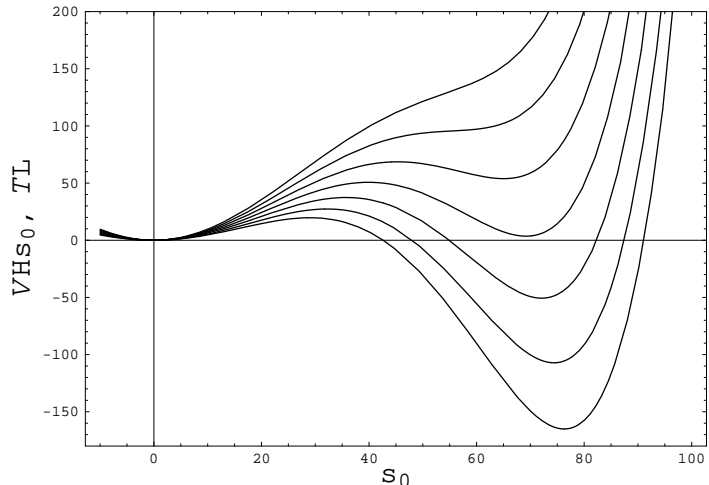}
\caption{Evolution of the effective potential
$V(\sigma_0,T)$ as a function of the order parameter $\sigma_0$
for several temperature steps: $T$ = 197, 193, 189, 185, 181, 177,
173 MeV from top to bottom. Two minima of $V(\sigma_0,T$ appear as
degenerate at $T_c \approx 185$ MeV. \label{fig:Fig3}}
\end{center} 
\end{figure}
Next let us compute numerically the effective potential
$V(\sigma_0,T)$, given by (\ref{3.4}), as a function of the
temperature and the order parameter. This will give more insight
into the nature of phase transition. In Fig.~\ref{fig:Fig3} is
depicted the evolution of $V(\sigma_0,T)$ against $\sigma_0$ for
several temperature steps. It is found that the two minima of
$V(\sigma_0,T)$ appear as degenerate at $T_{c2} \approx 185$ MeV.
For $T \gtrsim 193$ MeV, $V(\sigma_0,T)$ has only one minimum at
$\sigma_0 =0$. The shape of the potential confirms that a first
order phase transition occurs.

\subsection{\large \bf Hartree-Fock approximation in the broken symmetry case $c\neq 0$}

For completeness let us now consider the case $c \neq 0$, hence
the gap equation for the order parameter (\ref{2.14}) turns out to
be
\begin{eqnarray}
\left[ m^2 +\frac{\lambda}6 \sigma_0^2 +\frac{\lambda}2
P_f(M_\sigma) +\frac{\lambda}2 P_f(M_\pi)\right]\sigma_0 - c = 0,
\label{3.7}
\end{eqnarray}
and the SD equations (\ref{2.12}), (\ref{2.13}) remain unchanged.

\begin{figure}[h]
\begin{center}
\includegraphics[width=1\columnwidth]{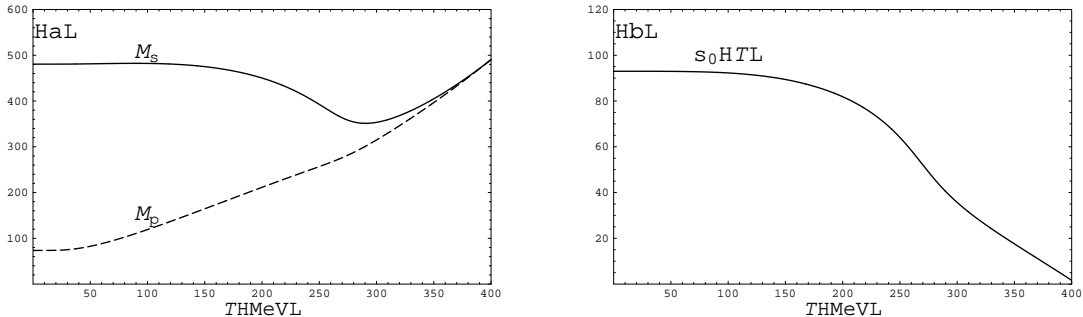}
\vskip .5cm \caption{$(a)$ Solution of the system of gap equations
for $c\neq 0$. At low temperature $M_\pi$ coincides with the
observed $m_\pi$. $(b)$ Evolution of the order parameter as a
function of temperature. \label{fig:Fig4}}
\end{center} 
\end{figure}

After solving numerically the system (\ref{2.12}), (\ref{2.13})
and (\ref{3.7}) we get the solution presented in
Fig.~\ref{fig:Fig4}.

\vspace{.5cm}

Comparing the above presented result with those obtained in
[16-19] we have an excellent agreement which shows that the
quantum effect is an negligible quantity.

\section{\large \bf THE HIGHER-LOOP DIAGRAMS CONTRIBUTION }
\label{sec:Hloop}

\begin{figure}[h]
\begin{center}
\begin{picture}(60,30)(0,0)
\GCirc(30,0){30}{1} \Line(0,0)(60,0) \Vertex(0,0){1.5}
\Vertex(60,0){1.5}
\end{picture} \vspace{1.5cm}
\caption{The sunset graph. \label{fig:Fig5}}
\end{center}
\end{figure}
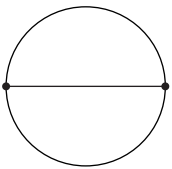

In the preceding section it was shown that the phase transition is
of first order. The Gaussian approximation \cite{prd32} in 3+1
dimensions gives the same result. However, the renormalization
group approach applied to the linear sigma model [25, 26]
indicates that the phase transition is of second order.  Arnold
and Espinosa \cite{prd47b} pointed out that loop diagrams other
than superdaisy ones are important near critical temperature. The
nature of the phase transition in the $\phi^4$ field theory
remains a basic question to be settled \cite{hepph99}. Therefore,
this requires a further investigation on higher-loop effect. In
this respect, it is necessary to incorporate the higher-loop
diagrams into consideration. The higher-loop graph next to the
double bubble one is the sunset graph given in Fig~\ref{fig:Fig5}

In a recent publication \cite{hepth00} Bordag and Skalozub
studied, in addition to the double bubble, an infinite series of
graphs shown in Fig~\ref{fig:Fig7}

\vspace{.5cm}
\begin{figure}[h]
\begin{center}
\begin{eqnarray*}
W_2^{(h)} = \frac{1}{12}\;
    \begin{picture}(40,20)(0,-5)
    \BCirc(20,0){20} \Line(20,20)(20,-20)  \Vertex(20,20){1} \Vertex(20,-20){1}
    \end{picture} + \frac{1}{48}\;
    \begin{picture}(40,20)(0,-5)
    \BCirc(20,0){20} \CArc(0,0)(29,-45,45) \CArc(40,0)(29,135,225)
    \Vertex(20,20){1} \Vertex(20,-20){1}
    \end{picture} +\sum_{n\geq 3}\frac{1}{2^n}\;
    \begin{picture}(40,20)(0,-5)
    \Oval(5,0)(20,5)(360) \Oval(35,0)(20,5)(360) \Line(5,-20)(35,-20)
    \Vertex(5,-20){1} \Vertex(35,-20){1} \Vertex(5,20){1} \Vertex(35,20){1}
    \GCirc(12,20){1.5}{.5} \GCirc(20,20){1.5}{.5}
    \GCirc(28,20){1.5}{.5}
    \end{picture} +\sum_{n\geq 3}\frac{1}{2^{n+1}n}\;
    \begin{picture}(40,20)(0,-5)
    \Oval(5,0)(20,5)(360) \Oval(35,0)(20,5)(360) \Oval(20,-20)(3,15)(360)
    \Vertex(5,-20){1} \Vertex(35,-20){1} \Vertex(5,20){1} \Vertex(35,20){1}
    \GCirc(12,20){1.5}{.5} \GCirc(20,20){1.5}{.5}
    \GCirc(28,20){1.5}{.5}
    \end{picture}
\end{eqnarray*} 
\caption{The big dots symbolize further insertion of the subgraph
$-\!\!\bigcirc\!\! -$ so that the number of vertices is $n$. The
3-particle vertex $\sim -\lambda \phi_0$, the 4-particle vertex
$\sim -\lambda$. \label{fig:Fig7}}
\end{center}
\end{figure}
which is motivated by the $1/N$ expansion.

Making use of the ansatz
\[
G^{-1}(p) = p^2 +M^2
\]
at high temperature they concluded that the contribution of this
set of graphs does not influence significantly on the physical
process, making the transition a bit stronger first order.

Next let us generalize the consideration including an infinite
series of diagrams depicted in Fig~\ref{fig:Fig8}

\vspace{.5cm}
\begin{figure}[h]
\begin{center}
\begin{eqnarray*}
    \begin{picture}(40,20)(0,-5)
    \BCirc(20,0){20} \Line(20,20)(20,-20)  \Vertex(20,20){1} \Vertex(20,-20){1}
    \end{picture} +
    \begin{picture}(40,20)(0,-5)
    \BCirc(20,0){20} \CArc(0,0)(29,-45,45) \CArc(40,0)(29,135,225)
    \Vertex(20,20){1} \Vertex(20,-20){1}
    \end{picture} +\!\!\sum_{n\geq 3}a_n\;
    \begin{picture}(40,20)(0,-5)
    \Oval(5,0)(20,5)(360) \Oval(35,0)(20,5)(360) \Line(5,-20)(35,-20)
    \Vertex(5,-20){1} \Vertex(35,-20){1} \Vertex(5,20){1} \Vertex(35,20){1}
    \GCirc(12,20){1.5}{.5} \GCirc(20,20){1.5}{.5}
    \GCirc(28,20){1.5}{.5}
    \end{picture} +\!\!\sum_{n\geq 3}b_n\;
    \begin{picture}(40,20)(0,-5)
    \GCirc(20,0){20}{1} \Line(20,-20)(20,-10)
    \Vertex(20,20){1} \Vertex(20,-20){1}
    \GCirc(20,-10){1.5}{.5} \Vertex(20,-2){.5}\Vertex(20,0){.5}\Vertex(20,2){.5}
    \GCirc(20,5){1.5}{.5} \GCirc(20,10){1.5}{.5} \GCirc(20,15){1.5}{.5}
    \end{picture}+\!\!\sum_{n\geq 3}c_n\;
    \begin{picture}(40,20)(0,-5)
    \Oval(5,0)(20,5)(360) \Oval(35,0)(20,5)(360) \Oval(20,-20)(3,15)(360)
    \Vertex(5,-20){1} \Vertex(35,-20){1} \Vertex(5,20){1} \Vertex(35,20){1}
    \GCirc(12,20){1.5}{.5} \GCirc(20,20){1.5}{.5}
    \GCirc(28,20){1.5}{.5}
    \end{picture}+\!\!\sum_{n\geq 3}d_n\;
    \begin{picture}(40,20)(0,-5)
    \GCirc(20,0){20}{1} \Vertex(20,20){1} \Vertex(20,-20){1}
    \GCirc(20,-15){1.5}{.5}
    \GCirc(20,-10){1.5}{.5} \Vertex(20,-2){.5}\Vertex(20,0){.5}\Vertex(20,2){.5}
    \GCirc(20,5){1.5}{.5} \GCirc(20,10){1.5}{.5} \GCirc(20,15){1.5}{.5}
    \end{picture}
\end{eqnarray*} 
\caption{The big dots symbolize further insertion of the subgraph
$-\!\!\bigcirc\!\! -$ so that the number of vertices is $n$.
\label{fig:Fig8}}
\end{center}
\end{figure}

Then the effective potential, corresponding to Fig~\ref{fig:Fig8},
reads
\[
V_2^{[8]} = \frac{3}{2} \lambda^2 \phi_0^2 H + \frac{\lambda^2}{2}
K,
\]
where
\begin{eqnarray*}
H &=& \mbox{Tr}_q G(q)\Sigma(q)\left\{
f_1[\lambda\Sigma(q)]+f_2[\lambda\Sigma(q)]\right\}, \\
K &=& \mbox{Tr}_q \Sigma^2(q)\left\{
g_1[\lambda\Sigma(q)]+g_2[\lambda\Sigma(q)]\right\}, \\
\Sigma(p) &=& \mbox{Tr}_q G(q)G(p+q) = -\!\!\bigcirc\!\! -
\end{eqnarray*}
and $ f_1, f_2, g_1, g_2 $ are respectively the functions
representing the series expansions given in Fig~\ref{fig:Fig8}
with the coefficients $ a_n, b_n, c_n, d_n$.

It is evident that
\begin{eqnarray*}
f_i[0] = g_i[0] = 1,\;\; i=1,2.
\end{eqnarray*}

Now it is easily seen that we arrive at the same conclusion by
applying the discussion of \cite{hepth00} to every fixed
expression for every function among $f_1, f_2, g_1$ and $g_2$.

\section{\large \bf CONCLUSION AND DISCUSSION}
\label{sec:Conc}

We have considered the $O(4)$ linear sigma model at finite
temperature within the framework of the CJT formalism restricted
to the double bubble diagram approximation. The renormalization of
the CJT effective potential for this model has been discussed in
many papers [14-19] and one succeeds only in the large-$N$ limit
case \cite{prd10a}, where all divergent terms are absorbed into
the bare quantities. For other case, however, not all divergences
can be removed away \cite{hepph97}. In the present paper,
following exactly the spirit of renormalization that requires all
divergent terms must be absorbed into counterterms, corresponding
to renormalizing mass and coupling constant, we impose two
constraints (\ref{2.9}) and (\ref{2.10}), to ensure that only
finite terms would be still present in the effective potential. It
is possible that this renormalization method is most suited for
renormalizing the CJT effective potential in loop approximations.
In order for the Goldstone theorem not to be violated the physical
masses of mesons are defined by (\ref{2.16}), (\ref{2.17}) and
propagators are chosen in the symmetric form (\ref{2.7}). The
solution of the gap equations, derived from the renormalized
effective potential, and the shape of the effective potential, as
a function of the order parameter exhibits a first order phase
transition with the critical temperature $T_c \simeq 185$ MeV.

To go further to higher-loop graphs we have shown that the order
of the phase transition is unchanged by incorporating an infinite
series of diagrams, in which the sunset one is a special case.

In summary, it is possible to confirm that the phase transition in
the $\phi^4$ field theory is of first order.

\section*{Acknowledgments}

\noindent The financial support from the Vietnam National Science
Foundation is acknowledged.


\end{document}